\documentclass[]{article}
\usepackage[utf8]{inputenc}
\usepackage{amsmath}
\usepackage[pdftex]{graphicx}
\usepackage{amscd}
\usepackage{calrsfs}
\usepackage{slashed}
\usepackage{amsfonts}
\usepackage{authblk}
\usepackage{caption}
\usepackage{subcaption}
\usepackage[letterpaper, portrait, margin=0.7in]{geometry}
\graphicspath{{figs/}}
\title{Black holes and the third law of thermodynamics revisited}
\author{Miguel Socolovsky}
\affil{Instituto de Ciencias Nucleares, Universidad Nacional Aut\'onoma de M\'exico, Cd. Universitaria, 04510, Ciudad de M\'exico, M\'exico\\socolovs@nucleares.unam.mx}

\providecommand{\keywords}[1]
{
	\small	
	\textbf{Keywords:} #1
}
\begin{document}
\date{}
\maketitle

\begin{abstract}
Black holes contradict the Nernst-Planck ($N/P$) version of the 3rd. law of thermodynamics, but agree with its unattainability (U) version. This happens without contradiction because the $N/P$ and $U$ versions are not equivalent, namely, $N/P$ implies $U$ but $U$ does not imply $N/P$. So, black holes obey the weaker version of the 3rd. law, but not the stronger one.
\end{abstract}

\

\section{Introduction}

\

It is commonly believed that the Nernst-Planck ($N/P$) version of the 3rd. law of thermodynamics and the unattainability ($U$) version are equivalent [1]. Nernst ($N$) version [2] asserts that in the $T\to 0_+$ limit of the absolute temperature, the entropy $S$ of the system tends to a constant which is independent of the remaining thermodynamic quantities that characterize the system (pressure, volume, magnetic field, etc.), while $N/P$ says that this constant is zero [3]. On the other hand, the $U$ version says that to reach $T=0$ needs an infinite amount of time or, what is equivalent, an infinite number of steps. 

\

In Section 2 we show, with two examples, how $N/P$ $\Rightarrow$ $U$; moreover, the left hand side of the implication needs to include the 1st. and the 2nd. laws of thermodynamics. It is clear that the above amounts to -$U$ $\Rightarrow$ -$N/P$ but does not imply that $U$ $\Rightarrow$ $N/P$ [4]. That is, the $N$ (or $N/P$) version is {\it stronger} than the $U$ version or, in other words, unattainability can hold even if $N/P$ does not.  

\

The considerations for the Schwarzschild and Kerr black holes are reserved to Sections 3 and 4. In Section 3 the thermodynamics of the Schwarzschild black hole immediately illustrates the violation of $N$ (or $N/P$) and simultaneously the fulfillment of $U$ [5]. For the more involved case of the Kerr black hole (Section 4), the study of the entropy-temperature diagram clearly shows the violation of the $N$ (or $N/P$) version, while the loss of analiticity of the entropy as a function of energy (mass) and angular momentum at $T=0$ indicates the presence of a phase transition into a naked singularity, and therefore the disappearance of the black hole itself at this temperature. That is, as a black hole, the system never attains $T=0$. These arguments can be considered as a complement to the rigorous proof by Israel [6] and the precisions of Wreszinski and Abdalla [7].

\

\section{N/P $\Rightarrow$ U}
Through the use of two kinds of systems, one hydrostatic and the other magnetic, we show how, by the well known zig-zag processes, the $N$ and obviously also the $N/P$ versions of the 3rd. law together with the 1st. and 2nd. laws, imply the unattainability version of the 3rd. law.

\

2.1 Hydrostatic system

\

Consider the picture in Fig. 1: each curve represents the entropy $S$ of the system as a function of $T$ at distinct values of pressure $p$, $p_1$ and $p_2$ with $p_2>p_1$, with the property that, as $T\to 0_+$, both $S(T,p_1)$ and $S(T,p_2)$ converge to $S_0$. 

\
\begin{figure}[h]
	\centering
	\includegraphics[width=.3\linewidth]{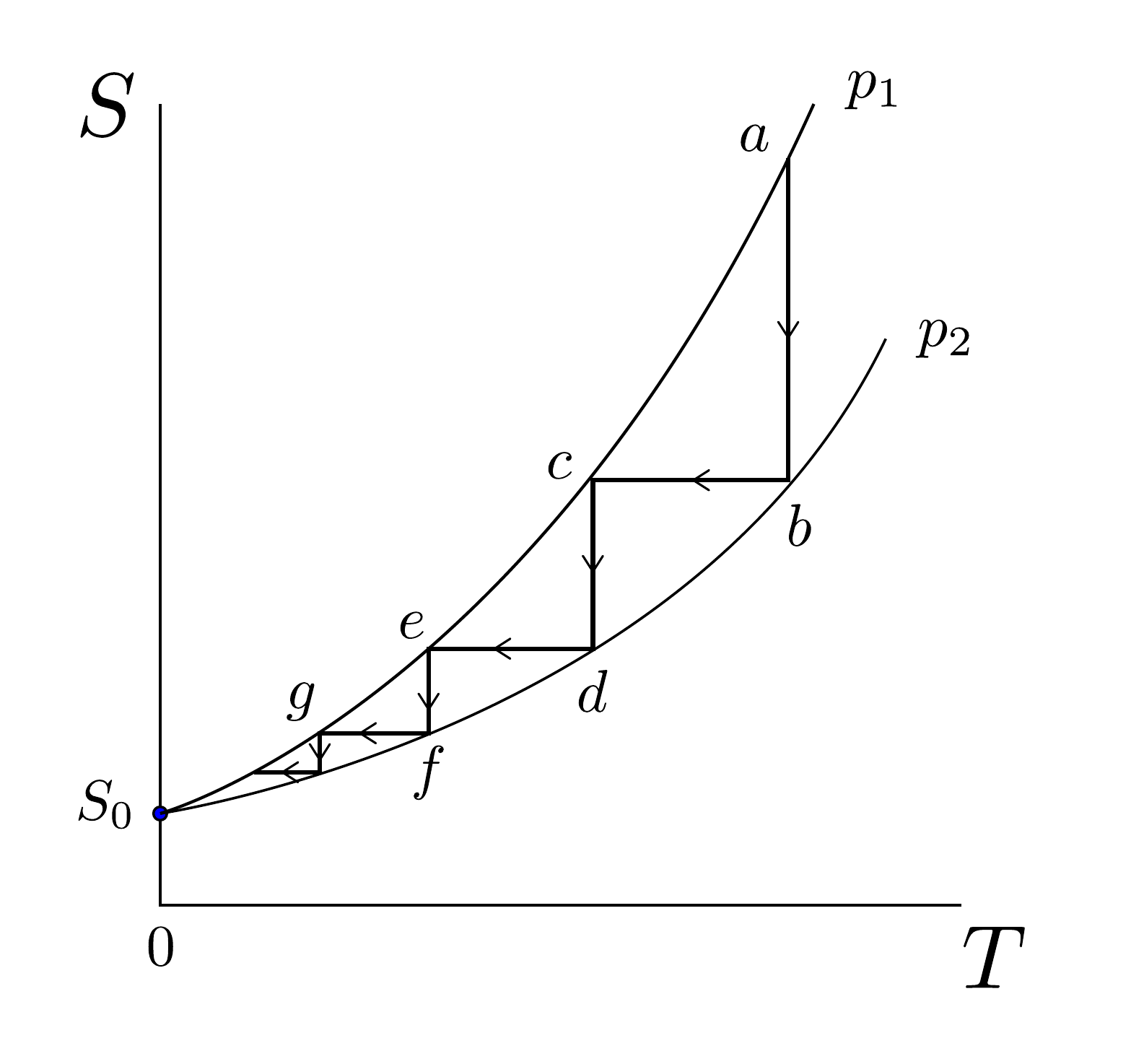}
	\caption{Zig-zag isothermal-adiabatic cooling for $S\to S_0$ as $T\to 0_+$}
\end{figure}

\

$a\to b$, $c\to d$, $e\to f$,..., are {\it isothermal compresions} which, from the  ``TdS" equation (consequence of the 2nd. law) $T\Delta S=C_p\Delta T-\alpha TV\Delta p$ [8], where $\alpha$ is the thermal expansion coefficient and $V$ the volume, reduce to $T\Delta S=-\alpha TV\Delta p$; since $\alpha$ and $V$ are positive, $\Delta p=p_2-p_1>0$ implies $\Delta S=S_b-S_a \ (S_d-S_c, \ S_f-S_e,...)<0$ i.e. a lowering of the entropy. The other part of the zig-zag's, $b\to c$, $d\to e$, $f\to g$,..., are {\it adiabatic expansions} ($\Delta V>0$): from the 1st. law the variation of the internal energy $\Delta U=T\Delta S-p\delta V$ reduces to $\Delta U=-p\Delta V<0$ which implies a lowering of $T$. It is clear that through this procedure an infinite quatity of each time smaller zig-zag's steps is needed to arrive at $T=0_+$. $\Box$

\

At the same time it is clear that a cooling to $T=0$ is possible in a finite number of zig-zag's if $S(0_+,p_1)\neq S(0_+,p_2)$ i.e. if $N$ does not hold. 

\

2.2 Magnetic system (paramagnetism)

\

Consider a system of $N$ spins 1/2, each with magnetic moment $\mu$ in the presence of an external magnetic field $B$. The picture of the entropy $S$ as a function of $T$, $B$, and $N$ is analogous to that in Fig. 1 with magnetic fields $B_1$ and $B_2$ respectively replacing $p_1$ and $p_2$ ($B_2>B_1$). $S$ is given by $S(T,B,N)=N(ln(2Chx)-xThx)$  where $x={{\mu B}\over{T}}$ [9]. The derivation of this entropy involves the 1st. and 2nd. laws of thermodynamics as well the canonical ensamble of equilibrium statistical mechanics. In this case $S_0=S(0_+,B,N)=0$. The vertical parts of the zig-zag's ($a\to b$,...) are magnetizing isothermals ($\Delta B>0$), while the horizontal parts ($b\to c$,...) are demagnetizing adiabatics ($\Delta B<0$). It is easy to verify that in each isothermal, $\Delta S\vert _{T,\Delta B>0}=-{{\mu}\over{T}}({{1}\over{2}}Th({{\mu B}\over{T}})+{{{{\mu B}\over{T}}\over{Ch^2h({{\mu B}\over{T}})}}})\Delta B<0$ (entropy descends), while in each adiabatic, $\Delta T\vert _{S,\Delta B<0}=-T{{\vert \Delta B\vert }\over{B}}<0$ (temperature descends). Again, an infinite quantity of each time smaller zig-zag's is needed to arrive at $T=0_+$. $\Box$ 

\

\section{Schwarzschild black hole}
It is well known that for the Schwarzschild black hole of mass $M$ and horizon radius $2M$, the entropy $S={{A}\over{4}}$ and the Hawking temperature $T={{\kappa}\over{2\pi}}$, where $A$ is the horizon area and $\kappa$ is the surface gravity, are given by 
\begin{equation}
	T={{1}\over{8\pi M}}\equiv T_{Schw.}
\end{equation}
and
\begin{equation}
	S=4\pi M^2\equiv S_{Schw.}
\end{equation}
respectively. So, for $T\to 0_+$, $M\to +\infty$ and therefore $S\to \infty$ at the absolute zero. The last result implies the violation of $N$ or $N/P$, and at the same time the fulfillment of $U$, due to the impossibility for a black hole to reach an infinite amount of mass or energy in any finite time, let it be proper or measured at $r=\infty$ [10]. 

\

\section{Kerr black hole}
For a Kerr black hole of mass $M$ and angular momentum $J$ ($0\leq J<M^2$) in Boyer-Lindquist coordinates, the temperature and entropy at the event horizon $r_+=M+\sqrt{M^2-({{J}\over{M}})^2}$ are respectively given by [11]
\begin{equation}
T(M,J)={{\sqrt{1-(J/M^2)^2}}\over{4\pi M(1+{\sqrt{1-(J/M^2)^2})}}}
\end{equation}
and
\begin{equation}
S(M,J)=2\pi M^2(1+\sqrt{1-(J/M^2)^2}) \ .
\end{equation}
At $J=0$ both quantities are continuous ($C^0$)	and reproduce the Schwarzschild values $T(M,0)=T_{Schw.}$ and $S(M,0)=S_{Schw.}$. Since
\begin{equation}
({{\partial T}\over{\partial J}})_M(M,J)=-{{J}\over{4\pi M^5}}\times {{1}\over{(1+\sqrt{1-(J/M^2)^2})^2\sqrt{1-(J/M^2)^2}}}
\end{equation}
and
\begin{equation}
({{\partial S}\over{\partial J}})_M(M,J)=-{{2\pi J/M^2}\over{\sqrt{1-(J/M^2)^2}}},
\end{equation}
$T$ and $S$ have also continuous first derivatives ($C^1$) for $0\leq J<M^2$ with 
\begin{equation}
({{\partial T}\over{\partial J}})_M(M,0_+)=0_- \ \ and \ \ ({{\partial S}\over{\partial J}})_M(M,0_+)=0_-.
\end{equation}
At $J=M^2$ both the event horizon at $r_+$ and the Cauchy horizon at $r_-$ coincide:
\begin{equation}
r_+=r_-=M,
\end{equation} 
the black hole region disappears, formally reaching the so called ``extreme black hole", with 
\begin{equation}
T(M,M^2)=0 \ \ and \ \ S(M,M^2)=2\pi M^2={{S_{Schw.}}\over{2}}, 
\end{equation}
and the first derivatives in (5) and (6) diverge:
\begin{equation}
({{\partial T}\over{\partial J}})_M(M,(M^2)_-)=-\infty \ \ and \ \ ({{\partial S}\over{\partial J}})_M(M,(M^2)_-)=-\infty.
\end{equation}
In other words, in the interval $J\in [0,M^2]$, $T$ and $S$ belong to $C^0$ but not to $C^1$, with $-\infty <({{\partial T}\over{\partial J}})_M$, $({{\partial S}\over{\partial J}})_M<0$ for $0<J<M^2$ (see Fig. 2).

\

\begin{figure}[h]
	\centering
	\begin{subfigure}{.5\textwidth}
		\centering
		\includegraphics[width=.5\linewidth]{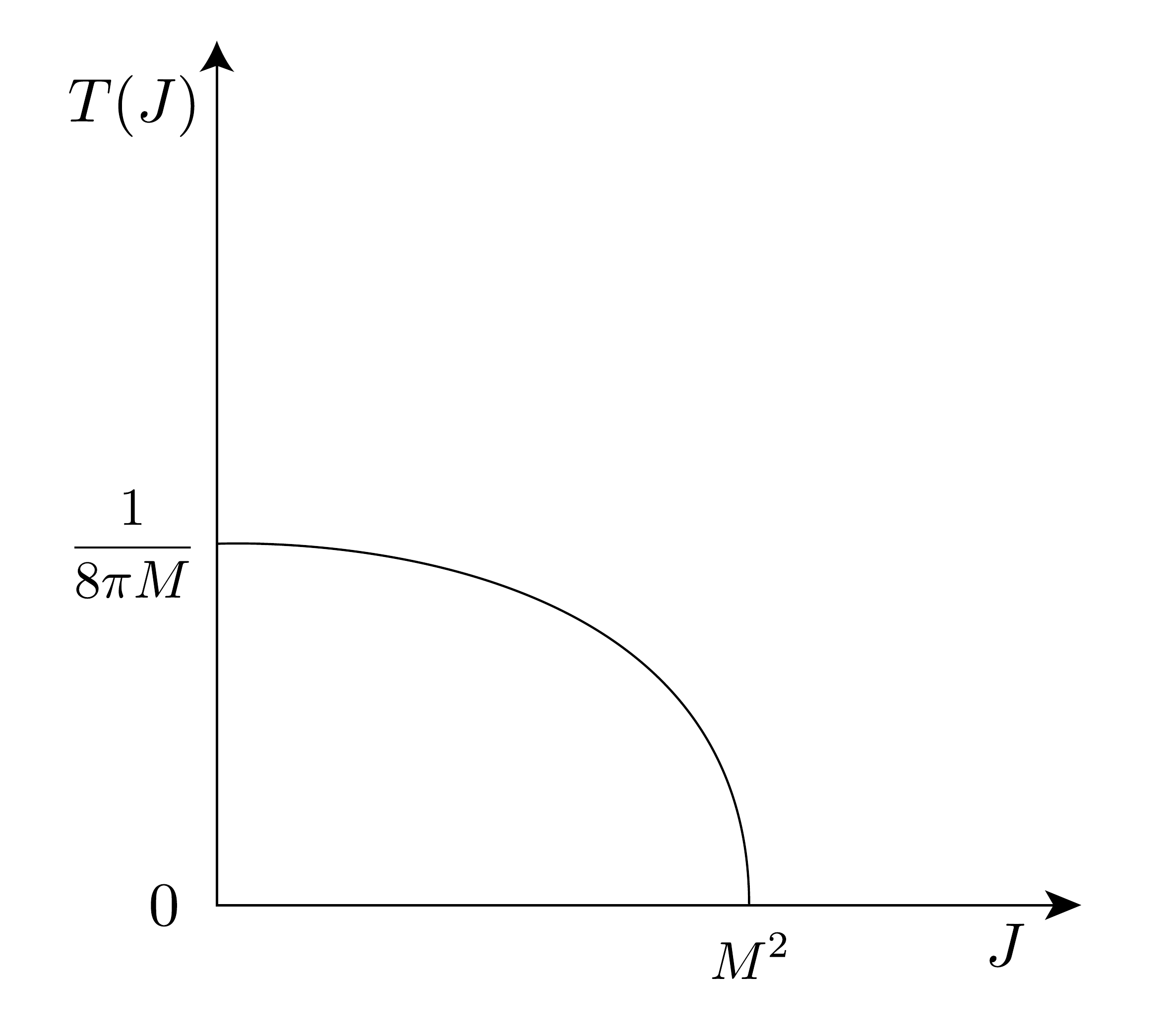}
		\label{fig:sub1}
	\end{subfigure}%
	\begin{subfigure}{.5\textwidth}
		\centering
		\includegraphics[width=.5\linewidth]{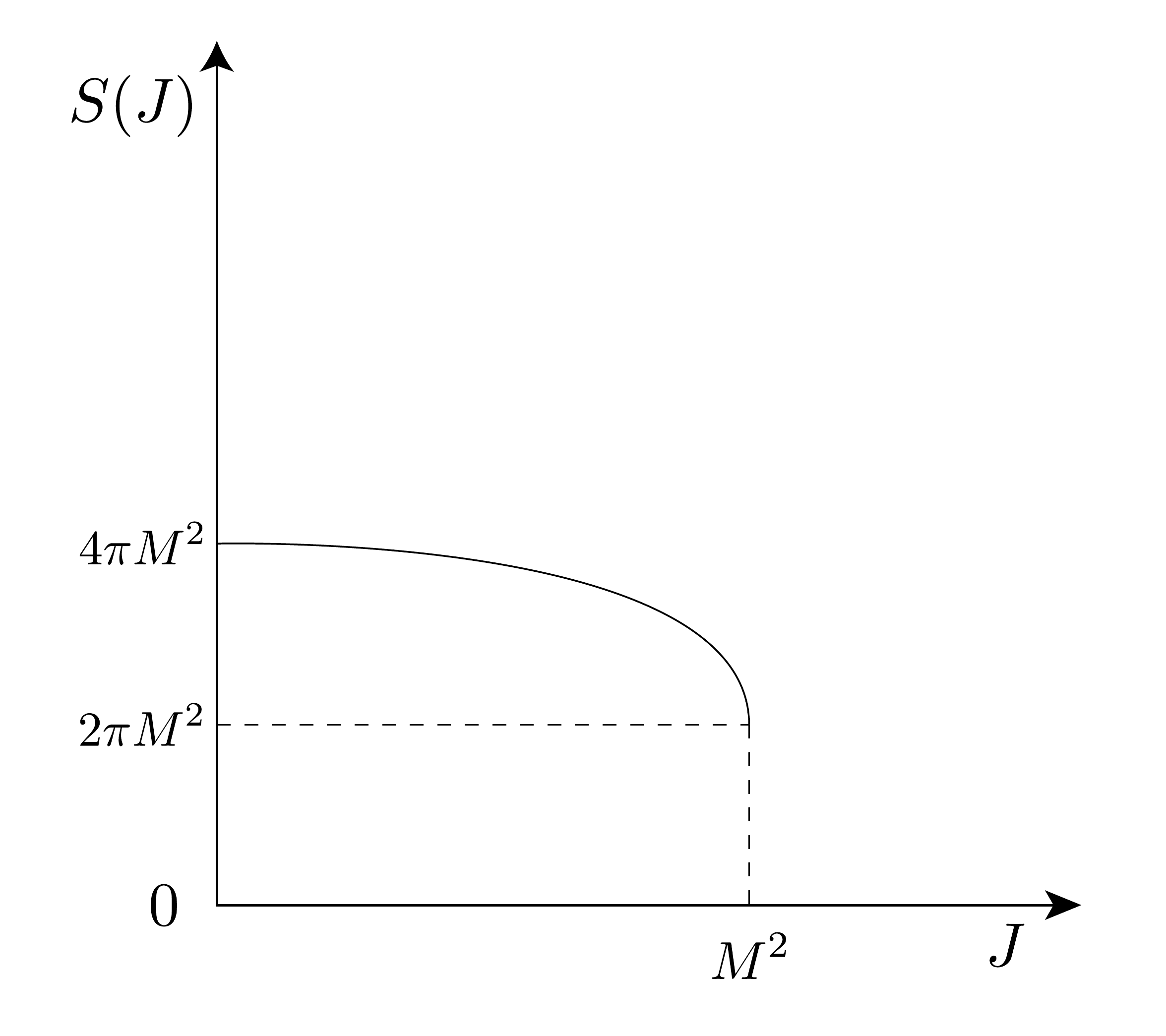}
		\label{fig:sub2}
	\end{subfigure}
	\caption{{Temperature and entropy as function of $J$ with $M$ fixed}}
\end{figure}

\

On the other hand,
\begin{equation}
({{\partial S}\over{\partial M}})_J(M,J)=4\pi M\times{{1+\sqrt{1-(J/M^2)^2}}\over{\sqrt{1-(J/M^2)^2}}}={{1}\over{T(M,J)}},
\end{equation}
which is $C^0$ and $C^1$ at $J=0$ since $({{\partial S}\over{\partial M}})_J(M,0)=8\pi M$ and 
\begin{equation}
{{\partial}\over{\partial J}}(({{\partial S}\over{\partial M}})_J)_M(M,J)={{4\pi J}\over{M^3}}\times{{1}\over{(\sqrt{1-(J/M^2)^2})^3}}
\end{equation}
with ${{\partial}\over{\partial J}}(({{\partial S}\over{\partial M}})_J)_M(M,0)=0$, but divergent at $J=(M^2)_-$ with $({{\partial S}\over{\partial M}})_J(M,(M^2)_-)=+\infty$, and therefore neither $C^0$ nor $C^1$ since ${{\partial}\over{\partial J}}({{1}\over{T(M,J)}})_M(M,(M^2)_-)=+\infty$ (see Fig. 3).

\
\begin{figure}[h]
	\centering
	\includegraphics[width=.3\linewidth]{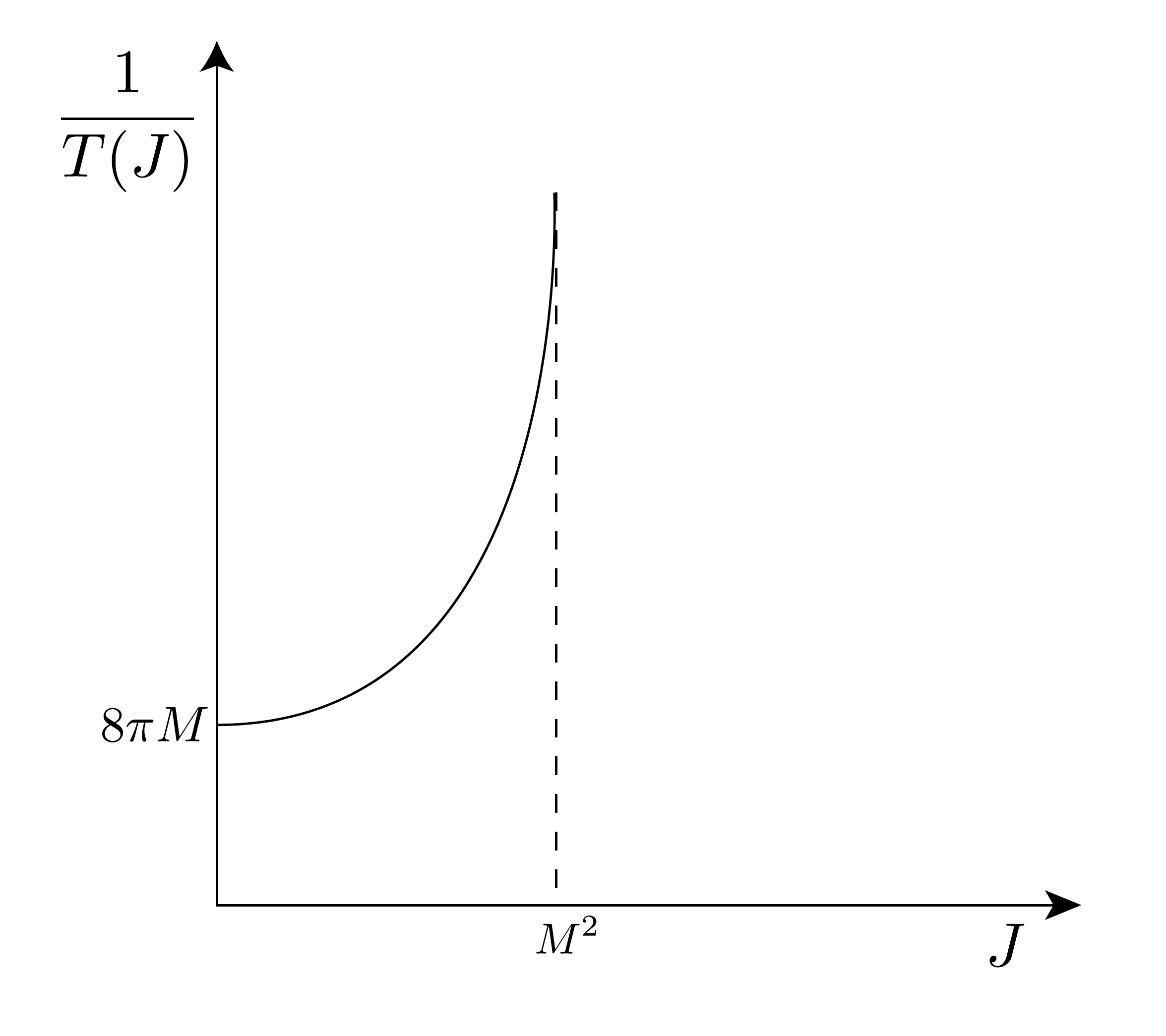}
	\caption{Inverse of absolute temperature as a function of $J$ for fixed $M$}
\end{figure}

\

I.e., $T(M,J)$ is monotonously decreasing with $J$ from ${{1}\over{8\pi M}}$ at $J=0$, to $0_+$ at $J=(M^2)_-$, at constant $M$.

\

It can be easily verified that
\begin{equation}
{{\partial}\over{\partial J}}(({{\partial S}\over{\partial M}})_J)_M(M,J)={{\partial}\over{\partial M}}(({{\partial S}\over{\partial J}})_M)_J(M,J)={{4\pi J}\over{M^3}}\times{{1}\over{(\sqrt{1-(J/M^2)^2})^3}}
\end{equation}
holds for all $J\in [0,M^2)$, while at $J=M^2$,
\begin{equation}
{{\partial^2S}\over{\partial J\partial M}}\vert_{J=(M^2)_-}={{\partial^2S}\over{\partial M\partial J}}\vert_{J=(M^2)_-}=+\infty.
\end{equation} 
The breaking down of the Maxwell-type relation ${{\partial^2S}\over{\partial J\partial M}}={{\partial^2S}\over{\partial M\partial J}}$ and therefore the analiticity of $S$ as a function of $(M,J)$ at $J=M^2$, is the indication of a {\it phase transition} into a naked singularity ocurring at $T=0_+$ and therefore of the unattainability ($U$) of this value of the absolute temperature. At the same time, the $M$-dependent value of the entropy at $T=0$ (or $T=0_+$) given by (9), shows that the $N$ or $N/P$ version of the 3rd. law is violated. 

\

As a function of $T$ at fixed $M$, $J$ is given by $J(T)=M^2{{\sqrt{1-8\pi MT}}\over{1-4\pi MT}}$.

\

Finally, we study $S$ as a function of $T$ for fixed $M$. From (3) and (4) 
\begin{equation}
S(T)={{2\pi M^2}\over{1-4\pi MT}}, \ \ 0<T\leq {{1}\over{8\pi M}}
\end{equation}  
with $S(0_+)=2\pi M^2$, $S({{1}\over{8\pi M}})=4\pi M^2$, and 
\begin{equation}
({{\partial S}\over{\partial T}})_M(T)={{8\pi^2M^3}\over{(1-4\pi MT)^2}}
\end{equation}
which equals $8\pi^2M^3$ at $T=0_+$ and $16\pi^2M^3$ at $T={{1}\over{8\pi M}}$ (see Fig. 4).

\

\begin{figure}[h]
	\centering
	\includegraphics[width=.3\linewidth]{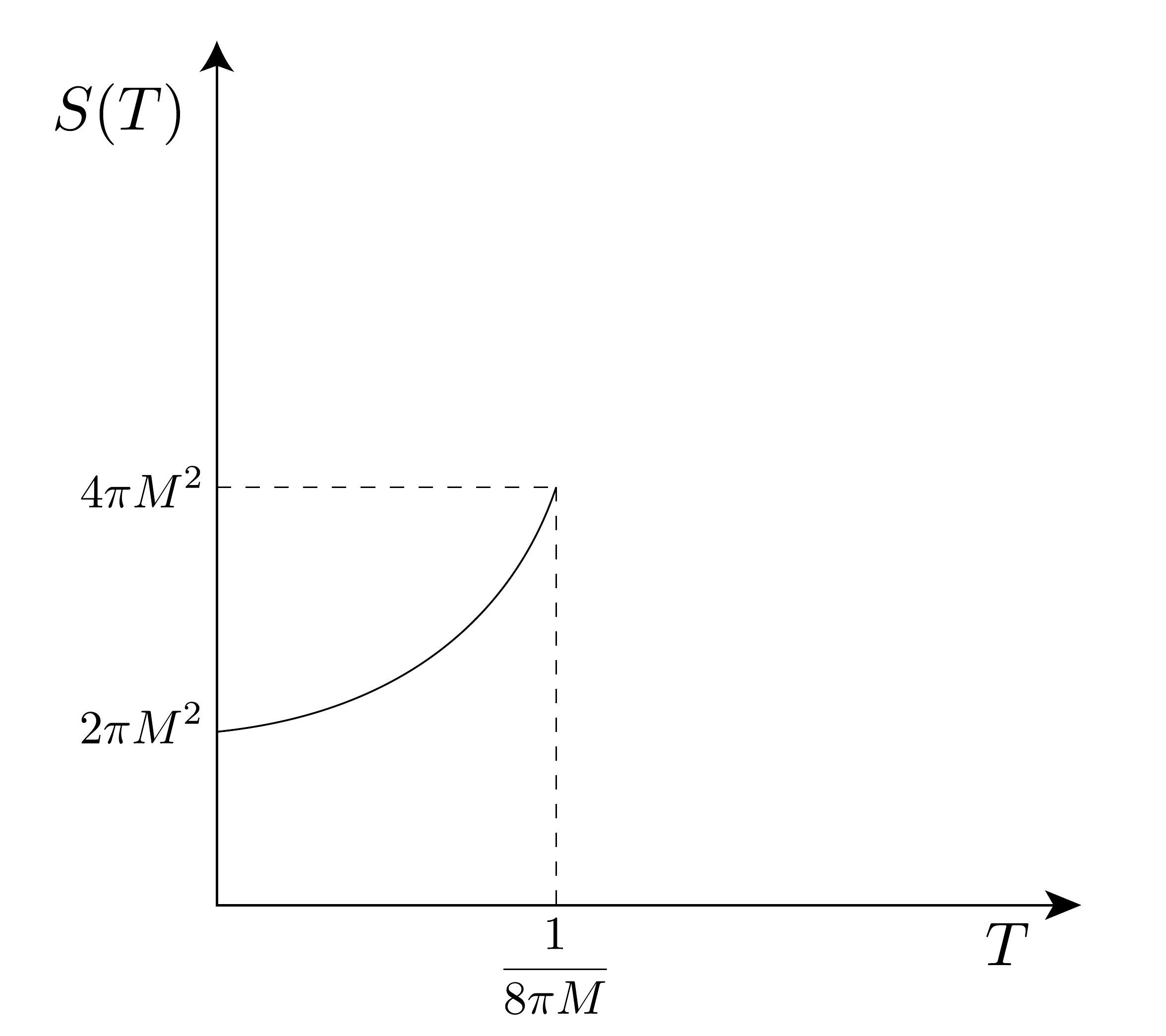}
	\caption{Entropy as a function of temperature for fixed $M$}
\end{figure}
\

{\bf Acknowledgments}

\

The author thanks for hospitality to the Instituto de Astronomía y Física del Espacio (IAFE) of the Universidad de Buenos Aires and CONICET, Argentina, where part of this work was done. Also, the author thanks Ernesto Eiroa at IAFE and Josué G.M. Trujillo at UASLP, México, for useful discussions, and Oscar Brauer at the University of Leeds, UK, for the drawing of the Figures.

\

{\bf References}\

\

[1] Belgiorno, F. and Martellini, M. (2004) Black Holes and the Third Law of Thermodynamics. {\it International Journal of Modern Physics D}, {\bf 13}, 739-770. https://doi.org/10.1142/S0218271804004876

\

[2] Nernst, W. (1906) Über die Berechnung Chemisher Gleichgewichte aus Termishen Messungen. {\it Nach. Kgl. Ges. Wiss. Gott.}, {\bf 1}, 1-40.

\

[3] Planck, M. (1911) {\it Thermodynamik}, 3rd. edn. (De Gruyter).

\

[4] Wheeler, J.C. (1991) Nonequivalence of the Nernst-Simon and Unattainability Statements of the Third Law of Thermodynamics. {\it Physical Review A}, {\bf 43}, 5289-5295. https://doi.org/PhysRevA.43.5289

\

[5] Wald, R.M. (1997) ``Nernst theorem" and Black Hole Thermodynamics. {\it Physical Review D}, {\bf 56}, 6467-6474.  

\
https://doi.org/10.1103/PhysRevD.56.6467

\

[6] Israel, W. (1986) Third Law of Black Hole Dynamics: A Formulation and Proof. {\it Physical Review Letters} {\bf 57}, 397-399. https://doi.org/10.1103/PhysRevLett.57.397

\

[7] Wreszinski, W.F. and Abdalla, E. (2009) A Precise Formulation of the Third Law of Thermodynamics. {\it Journal of Statistical Physics}, {\bf 134}, 781-792. https://doi.org/ 10.1007/s10995-009-9693-5

\

[8] Huang, K. (1963) {\it Statistical Mechanics}, J. Wiley, New York-London-Sydney; p. 22.

\

[9] Schmalian, J. (2012) {\it Lecture Notes, Statistical Mechanics (Theory F)}. Institute for Theory of Condensed Matter, Kalsruhe Institute of Technology; p. 25.

\

[10] Lüst, D. and Vleeshouwers, W. (2019) {\it Black Hole Information and Thermodynamics}, Springer; p. 47.

\

[11] Socolovsky, M. (2022) Kerr-Newman Black Holes and Negative Temperatures. {\it Journal of High Energy Physics, Gravitation and Cosmology}, {\bf 8}, 940-947. https://doi.org/10.4236/jhepgc.2022.84065

\end{document}